%
\overfullrule=0pt
\input phyzzx

\font\fourteendunh=cmdunh10 scaled\magstep2
\Pubnum={UCLA/90/TEP/72}
\date={}
\def\ucla{Department of Physics\break University 
of California Los Angeles\break
	Los Angeles, California 90024--1547}
\titlepage
\title{Quantum gravitational measure for three-geometries
\fourteendunh \foot
{\rm This work was supported in part by Department of Energy grant 
DE-AT03-88ER 40384}}\foot{\rm Published in Phys. Lett. {\bf B262}, 405 (1991).}
\author{Pawel O. Mazur}\foot{\rm E-mail address: mazur@psc.psc.sc.edu}
\foot{\rm Present address: Department of Physics, University of South 
Carolina, Columbia, SC 29208}
\address{\ucla}
\abstract{The gravitational measure on an arbitrary topological 
three-manifold is constructed. 
The nontrivial dependence of the measure 
on the conformal factor is discussed. 
We show that only in the case of a compact manifold with boundary 
the measure acquires a nontrivial dependence on the conformal factor 
which is given by the Liouville action. A nontrivial Jacobian 
(the divergent part of it) generates the Einstein-Hilbert action. 
The Hartle-Hawking wave function of Universe 
is given in terms of the Liouville action. In the 
gaussian approximation to the Wheeler-DeWitt equation this result 
was earlier derived by Banks et al. Possible connection 
with the Chern-Simons gravity is also discussed.

}
\endpage
%

General covariance plays the central role in physics. 
Quantization of systems with a large symmetry 
(general covariance, gauge invariance) 
requires proper geometrical tools. 
Indeed, the Feynman functional integral approach 
has proven extremely useful in formulating properly 
the quantum dynamics 
of random paths (a point particle) and surfaces 
(strings, two-geometries) [1,2,3]. 
The problem of great interest now is how to extend the Feynman 
and the Polyakov approach to the case of quantum geometries ($d>2$) 
[4]. It seems that we require the knowledge of how 
to sum over random three- and four-geometries in order 
to approach the problem of quantum gravity. 
The simpler case of two-geometries was 
understood in recent several years [2,3,5,]. 
Recently the subject of non-critical strings or induced 
quantum 2d gravity with its Liouville theory formulation 
has received great attention [5]. 
Also, recent interest in topology change and wormholes 
(topology changing amplitudes in quantum gravity) 
seems to indicate that the rigorous 
proper gravitational measure on manifolds 
of an arbitrary topology is required in order to make 
discussion of these issues more quantitative.
\par
In this Letter we study for simplicity the gravitational measure 
for three-geometries. This is an example of the general result for 
the quantum gravitational measure in any dimension. 
We extend the Polyakov approach to the case of 
three-geometries and show some qualitative differences. 
In particular, the absence of conformal anomalies 
in three dimensions (on three-geometries without boundary) 
makes a difference. However, the nontrivial Jacobian 
in the measure must be regularized and as a result 
we must add the bare local Einstein-Hilbert action as 
a local counterterm to the effective action. 
The local Einstein-Hilbert action is generated 
by the Jacobian in the measure. 
It seems that all infinities generated 
by the measure can be taken care off 
by simple renormalization of the cosmological 
and Newton constant in the Einstein-Hilbert action. 
In this sense we expect that the theory 
satisfy the criterium of calculability. 
\par
The subject of the rest of this Letter is the geometrical 
construction of gravitational measure 
in any dimension $d$ and calculation of the induced 
effective action for the conformal factor for three-geometries. 
\par
Consider the space of metrics $Riem(M)=Q(M)$ on a given manifold $M$. 
A tangent space $TQ_{|g}$ at a metric $g$ is spanned by the vectors 
${\delta}g$. The volume form on $Q$ can be introduced (almost) uniquely 
after $Q$ is equipped with the riemannian structure, i. e., a quadratic 
form $<,>_{TQ}$ on the tangent space $TQ$. 
It is the property of the Riemannian geometry that a metric defines 
the unique volume form ${\sqrt{g}}d^{n}x$. This is the ``square-root 
of determinant of the metric'' rule which determines the measure 
(volume form). The requirement of ultralocality of the measure 
leads to (an almost) unique metric on $Q$. This is the well known 
DeWitt metric on $Riem(M)$ [7] utilized by Polyakov in his work on 
quantum (random) two-geometries [2]
$$G^{abcd}={1\over 2}(g^{ac}g^{bd}+g^{ad}g^{bc})+Cg^{ab}g^{cd}, 
\eqno(1)$$
where $C>-{2\over d}$.
With this metric one defines the norm or scalar product on $TQ$ 
$$||{\delta}g||^2=<{\delta}g,{\delta}g>=
\int dx{\sqrt g}G^{abcd}\delta{g_{ab}}\delta{g_{cd}} . \eqno(2)$$
The Polyakov gaussian measure is defined by the condition 
$$\int d{\mu}({\delta}g)e^{-{1\over 2}||{\delta}g||^2}=1 . \eqno(3)$$
The space of metrics $Q$ has locally a form of the fiber bundle 
$Q={S\times Diff}$, where $S=Q/Diff$ is the Wheeler-DeWitt superspace 
defined as a space of metrics modulo diffeomorphisms. At some ``points'' 
$g_0$ of $Q$ which are fixed ``points'' of a subgroup of $Diff$ which is 
the isometry group of $g_0$, $Isom(g_0)$, $S$ is not a manifold but 
rather an infinite dimensional version of an orbifold [6]. 
In order to define a measure on $S$ we need to divide out the 
invariant measure on the fiber, the gauge group $Diff(M)$ of general 
covariance. We will proceed by introducing the natural co-ordinate 
system a part of which are group coordinates on $Diff(M)$. 
The space of metrics on a manifold $M$, $Q(M)=Riem(M)$ is a space of 
maps from $M$ into the coset space ${GL^{+}(d,R)/ SO(d)}$. 
Utilizing the decomposition of the coset ${GL^{+}(d,R)/SO(d)}=
{R^{+}\times {SL(d,R)/SO(d)}}$ we notice that the $R^{+}$ factor 
corresponds to an arbitrary positive scale of a metric which can be 
changed by the action of two groups of transformations. One of them is 
the group of diffeomorphisms $Diff_0(M)$ and the other is the Weyl 
conformal rescaling group $Weyl(M)$. Consider a conformal equivalence 
class of metrics on $M$, $g\equiv {\bar g}$ iff $g=e^{2\sigma}{\bar g}$, 
where $e^{2\sigma}$ is the nowhere vanishing positive function on $M$. 
${\bar g}$ will be called a representative element of the conformal 
equivalence class. This equivalence relation transforms $Q(M)=Riem(M)$ 
into the conformal geometry ${\it C(M)}={Riem(M)\over Weyl(M)}$. 
The Weyl group of conformal rescalings $Weyl(M)$ acts freely on 
${\it C(M)}$. An element (point) of $Riem(M)$, $g$ is left invariant 
under the transformation ${\bar g}\rightarrow e^{2\chi}{\bar g}$, 
${{\sigma}\rightarrow {\sigma}-{\chi}}$. It is important to notice 
that Weyl invariance corresponds to the freedom of choice of a 
representative element ${\bar g}$ of the conformal geometry ${\it C(M)}$. 
\par
The group of diffeomorphisms $Diff(M)$ acts on $Q(M)$ 
in the following way. 
Let $f$ is an element of $Diff(M)$, then $g^{f}=f^{*}g$, 
$$g_{a'b'}(x')={{\partial}x^c\over {\partial}x^{a'}}
{{\partial}x^d\over 
{\partial}x^{b'}}g_{cd}(x=f(x')), $$
$$x^a=f^a(x'), x^{a'}=(f^{-1})^a(x) . \eqno(4)$$
We can write this more concisely, $g^f={\omega}_f{\omega}_f g(f)$, 
where ${\omega}_f=df f^{-1}$ is the right-invariant one-form on 
$Diff(M)$. Once we choose the representative element ${\bar g}$ of the 
conformal geometry ${\it C(M)}$ we can parametrize 
the space of metrics in the following way
$$g=\left(e^{2\sigma} {\bar g}\right)^f .  \eqno(5)$$
Starting with an element ${\bar g}$ of the conformal 
class of metrics ${\it C(M)}$ the action 
of the semidirect product of $Diff(M)$ and 
$Weyl(M)$ produces a generic metric $g$ from $Q(M)=Riem(M)$. 
In this way we constructed a natural coordinate system on $Q(M)$, 
with $({\sigma},f)$ the natural group coordinates on 
$Weyl(M)$ and $Diff(M)$. 
The infinitesimal diffeomorphisms (connected to 
identity elements of $Diff_0(M)$) are generated 
by vector fields $\xi$. The integrated (finite) 
elements of $Diff_0(M)$ generated by $\xi$ are 
denoted by $Exp(\xi)$. The coordinate system on $Q(M)$ 
is given by a triple $(\sigma,\xi,{\bar g})$ 
which can be used in the construction of 
the gravitational measure. The change of variables 
$g\rightarrow(\sigma,\xi,{\bar g})$ requires a Jacobian 
which we evaluate below after the orthogonal decomposition 
of a tangent vector ${\delta}g\in TQ_{|g}$ 
with respect to the De Witt metric (1) is obtained. 
\par
The combined infinitesimal action of $Weyl(M)$ 
and $Diff(M)$ on $Q(M)$ produces an element ${\delta}g$ 
of the tangent space $TQ(M)$
$${\delta}g_{ab}=\left(2{\delta\sigma}+
{2\over d}{\nabla}_c{\xi}^c\right)g_{ab}+
{\left(L\xi\right)_{ab}} , \eqno(6)$$
where $L\xi$ is the trace-free part 
of the Lie derivative of a metric 
${\delta}_{f}g_{ab}={\nabla}_a{\xi}_b+{\nabla}_b{\xi}_a$, 
$$\left(L\xi\right)_{ab}={\nabla}_a{\xi}_b+
{\nabla}_b{\xi}_a-{2\over d}g_{ab}{\nabla}_c{\xi}^c .  
\eqno(7)$$
The operator $L$ maps vectors into trace-free 
symmetric tensors. The kernel of $L$, $KerL$, 
consists of the conformal Killing vectors (CKV). 
We can define a natural adjoint $L^{\dagger}$ 
of the operator $L$ with respect to 
the De Witt scalar product (2) on $TQ(M)$
$$<h,L\xi>=<L^{\dagger}h,\xi> , $$
$$\left(L^{\dagger}h\right)_a=-2{\nabla}^b{h_{ab}} ,  
g^{ab}h_{ab}=0 .  \eqno(8)$$
The adjoint operator $L^{\dagger}$ maps symmetric 
trace-free tensors into vectors. It is clear 
that these elements of the tangent space $TQ(M)$ 
which are not in the range of $L$, $Range(L)$, 
must be in the kernel of $L^{\dagger}$, $KerL^{\dagger}$. 
Those are the transverse trace-free symmetric tensors $h_{ab}$. 
Therefore, we have a natural orthogonal splitting 
of the vector space $TQ(M)_{|g}$
$${\delta}g_{ab}=\left(2{\delta\sigma}+
{2\over d}{\nabla}_c{\xi}^c\right)
g_{ab}+{\left(L\xi\right)}_{ab}+h_{ab} .  \eqno(9)$$
To find the measure $d{\mu}({\delta}g)={\it D}{\delta}g$ 
on the space of metrics we need to choose 
the representative element ${\bar g}\in{\it C}(M)$ 
of the conformal class of metrics ${\it C}(M)$. 
This can be done by specifying ${\bar g}$ 
to lie in a slice ${\bar Y}$ transversal 
to the orbits of $Weyl(M)$ and $Diff_0(M)$. 
Such a slice ${\bar Y}$ may be taken to be the Schoen-Yamabe 
slice defined by the condition of constant Ricci scalar: 
${\bar Y}=\lbrace {\bar g}, R({\bar g})=const \rbrace$. 
The recent proof of the Yamabe conjecture 
[8] for $d=3,4,5$ guarantees the local tranversality of ${\bar Y}$ 
to the orbits of $Weyl(M)$. 
For $d>2$ the Schoen-Yamabe slice ${\bar Y}$ 
is infinite dimensional; for $d=2$ its dimension 
is finite and any metric ${\bar g}$ can be 
parametrized by the Teichmuller (moduli) parameters $t_i$, 
${\bar g}={\bar g}(t)$. Formally, we can parametrize 
a metric ${\bar g}\in {\bar Y}$ by an infinite number 
of parameters $t_i$ (arbitrary functions), 
${\bar g}={\bar g}(t)$, and tangents to ${\bar Y}$ 
are symmetric tensors ${\bar {\phi}}_j$ 
defined by ${\delta}{\bar g}(t)=
\sum_j{\delta}t_j{\bar {\phi}}_j$ . 
The coordinate vectors along ${\bar Y}$ 
are ${\delta\sigma}{\bar g}$, ${\bar L}\xi$ 
and ${\bar {\phi}}_j$. Once the coordinate system 
on $Q(M)$ is specified we can easily 
evaluate the Jacobian at an arbitrary metric $g$ (5). 
Under the Weyl rescaling $e^{2\sigma}$ the tangent vectors 
to ${\bar Y}$ transform as: ${\phi}_j(e^{2\sigma}{\bar g})=
e^{2\sigma}{\bar {\phi}}_j({\bar g})$. 
This choice is a natural one for conformal geometry 
${\it C}(M)$ and it will be justified later. 
\par
It will be useful to introduce a basis of vectors 
$h_j$ in the kernel of $L^{\dagger}$, 
$h_j\in{KerL^{\dagger}}$. Using the orthogonal 
decomposition (9) of ${\delta}g_{ab}$ we obtain 
the following expression for the measure
$${\it D}{\delta}g=d{\mu_g}({\delta g})=
{det<{\phi}_j|h_k>_g\over 
det^{1/2}<h_j|h_k>_g}{\left(det'L^{\dagger}L\right)^{1/2}}
{\it D}{\sigma}{\it D}'{\xi}{\it D}t.\eqno(10)$$
In the expression for the measure all 
the determinants must be properly 
regularized because all of the matrices 
or operators act in the infinite dimensional spaces. 
The reason that determinants of matrices of scalar 
products appear is because the basis in $KerL^{\dagger}$ 
is not tangent to the Schoen-Yamabe slice ${\bar Y}$. 
We must consider the orthogonal projection 
of vectors in $T{\bar Y}$ onto $KerL^{\dagger}$. 
This way the generalized product 
of cosines of angles between $\phi_j$ and $h_k$ 
multiplies the standard measure on $T{\bar Y}$ 
(see, e.g., D'Hoker and Phong 1988) [3]. 
\par
The presence of matrices of scalar products 
$<\phi_j|h_k>$, $<h_j|h_k>$ makes it important 
to study their transformation laws under the Weyl 
rescaling. We assume that the kernel of $L^{\dagger}$, 
$KerL^{\dagger}$, is conformally invariant for any $d$. 
Indeed, one can easily see that the conformal weight 
$w$ of $h_j\in KerL^{\dagger}$ defined by 
$h_j=e^{w\sigma}{\bar h}_j$ is $w=2-d$ 
if $KerL^{\dagger}$ is conformally invariant
$${\left(L^{\dagger}h_j\right)_a}=e^{(w-2)\sigma}
\left[\left({\bar L}^{\dagger}{{\bar h}_j}\right)_a-
2(w+d-2){{\bar h}_{ja}}^b
{\nabla_b\sigma}\right] . \eqno(11)$$
The conformal variation of $L^{\dagger}$ can be easily 
read off from the last formula
$${\delta}L^{\dagger}=-2{\delta\sigma}L^{\dagger}-
2(d-2){\nabla\delta\sigma} . \eqno(12)$$
In a similar way we find the conformal 
weight $s$ of vectors in $KerL$. 
Consider the transformation 
$\xi_a=e^{s\sigma}{\bar {\xi}}_a$, then 
$$(L\xi)_{ab}=
e^{s\sigma}\left[({\bar L}{\bar {\xi}})_{ab}+
(s-2)\left(\nabla_a\sigma{\bar {\xi}}_b+
\nabla_b\sigma{\bar {\xi}}_a-
{2\over d}{\bar g}_{ab}
{\bar {\nabla}}^c{\bar {\xi}}_c\right)\right] . 
\eqno(13)$$
When $s=2$, then $KerL$ is conformally invariant. 
>From (13) we find also 
the conformal variation of $L$,
$$\delta(L\xi)_{ab}=
-2\left(\nabla_a\delta\sigma\xi_b+
\nabla_b\delta\sigma\xi_a-
{2\over d}g_{ab}\nabla^c\delta\sigma\xi_c\right) . 
\eqno(14)$$
The choice of conformal properties of vectors 
from $KerL^{\dagger}$ and 
$T{\bar Y}$ imply that $<\phi_j|h_k>_g=
<{\bar {\phi}}_j|{\bar h}_k>_{\bar g}$. 
This property of scalar products 
will prove to be quite important later. 
\par
The Jacobian $J$ in the gravitational measure, 
$J=\left(det'L^{\dagger}L\right)^{1/2}$ carries 
important information about how the measure changes 
along the orbits of the Weyl group of conformal 
rescalings $Weyl(M)$. As usual, we can find the dependence 
of $J$ on the conformal factor $\sigma$ 
by integrating the conformal anomaly equation. 
Of course, the situation is more subtle because 
of the possibility of zero modes of $L$, which are $CKV$. 
We denote $L^{\dagger}L=\Delta$. This is a vector operator 
mapping vectors into vectors which has the following form
$${\Delta_a}^b=-2\left({\nabla}^2{\delta_a}^b+(1-{2\over d})
{\nabla_a}{\nabla^b}+{R_a}^b\right) . \eqno(15)$$
We notice in passing that $\Delta$ has a diagonal symbol 
only for $d=2$, which makes the heat-kernel approach calculation 
of the conformal anomaly much easier in this case. 
For $d\neq 2$ the nondiagonal symbol of $\Delta$ 
complicates calculations significantly. 
\par
We evaluate the conformal change of $lnJ$ 
using the Schwinger proper time definition 
of the functional determinant 
$$lnJ=-{1\over 2}Tr'{\int_{\epsilon}}^{\infty}{dt\over t}
e^{-t\Delta} , \eqno(16)$$
where $Tr'$ denotes the functional trace operation 
over non-zero modes of $\Delta$. 
As usual, we calculate first a variation of $lnJ$ under 
the local change of the conformal factor $\sigma$
$$\delta{lnJ}={1\over 2}Tr'{\int_{\epsilon}}^{\infty}dt\delta\Delta 
e^{-t\Delta} . \eqno(17)$$
Using the operator identity $Ae^{-BA}=e^{-AB}A$, 
eqs. (12),(14) and the expression for the variation of $\Delta$, 
$\delta\Delta=\delta{L^\dagger}L+L^{\dagger}\delta L$, 
we obtain the following formulas
$$Tr'\delta\Delta e^{-t\Delta}=Tr'\left(\delta{L^\dagger}L
e^{-t{L^\dagger}L}\right)+Tr'\left({L^\dagger}\delta L
e^{-t{L^\dagger}L}\right) , \eqno(18)$$
$$Tr'\delta{L^\dagger}Le^{-t{L^\dagger}L}=
(d-2)Tr'\delta\sigma LL^{\dagger}
e^{-tLL^{\dagger}}-
dTr'\delta\sigma L^{\dagger}Le^{-tL^{\dagger}L} , 
\eqno(19)$$
$$Tr'L^{\dagger}\delta Le^{-tL^{\dagger}L}=
2Tr'\delta\sigma LL^{\dagger}
e^{-tLL^{\dagger}}
-2Tr'\delta\sigma L^{\dagger}Le^{-tL^{\dagger}L} . 
\eqno(20)$$
Collecting these results together we obtain 
a formula which allows for 
an explicit proper time integration in (17)
$$Tr'{\delta\Delta}e^{-t\Delta}=
(d+2)Tr'\delta\sigma{d\over dt}
e^{-tL^{\dagger}L}-dTr'\delta\sigma{d\over dt}e^{-tLL^{\dagger}} . 
\eqno(21)$$
Finally, we find the following useful expression 
for the conformal variation of $lndet'L^{\dagger}L$
$$\delta lndet'L^{\dagger}L=
dTr'\delta\sigma e^{-\epsilon LL^{\dagger}}-
(d+2)Tr'\delta\sigma e^{-\epsilon L^{\dagger}L} . \eqno(22)$$
We notice that in all these formulas the trace 
is taken over nonzero modes. 
Introducing projectors $P(L)$, $P(L^{\dagger})$ 
on the respective kernels, 
$KerL$ and $KerL^{\dagger}$, 
we can rewrite the last formula in the 
following way
$$\delta lndet'L^{\dagger}L=
dTr\delta\sigma e^{-\epsilon LL^{\dagger}}-
(d+2)Tr\delta\sigma e^{-\epsilon L^{\dagger}L}+
(d+2)Tr\delta\sigma P(L)
-dTr\delta\sigma P(L^{\dagger}) . \eqno(23)$$
At this point the previous discussion 
of conformal weight assignment for vectors 
from $KerL$ and $KerL^{\dagger}$ proves quite useful. 
We find
$$\delta lndet<\xi_j|\xi_k>=(d+2)Tr\delta\sigma P(L) , 
\eqno(24)$$
where $\xi_j\in KerL$ and 
$$\delta lndet<h_j|h_k>=-dTr\delta\sigma P(L^{\dagger}) . 
\eqno(25)$$
\par
The short time ($\epsilon\rightarrow 0^+$) 
expansion of heat kernels for both operators 
$L^{\dagger}L$ and $LL^{\dagger}$ is required for the 
calculation of a conformal anomaly 
for the vector operator $\Delta$. 
Integrating the conformal anomaly leads to 
a local effective action in the conformal gauge. 
We can see that the presence of zero modes in 
the Jacobian is universal for any dimension $d$
$$\eqalign{{\delta ln\left[{det'L^{\dagger}L
\over det<{\xi}_j|{\xi}_k>
det<h_j|h_k>}\right]^{1/2}}&
={1\over 2}\left[dTr{\delta\sigma} 
e^{-\epsilon LL^{\dagger}}-
(d+2)Tr{\delta\sigma} e^{-\epsilon L^{\dagger}L}
\right]\cr &=-\delta\Gamma ,\cr}\eqno(26)$$
where $\Gamma$ is the effective action for 
the conformal factor $\sigma$, 
which in the case of two-geometries was shown by Polyakov 
to be given in terms of the celebrated Liouville action. 
Our goal is to find the analog of the Liouville action 
in higher dimensions. The quantum dynamics of the conformal factor 
$\sigma$ in higher dimensions (especially for $d=4$) 
is governed by the renormalizable, but nonlinear, 
effective action $\Gamma[\sigma;{\bar g}]$. 
In order to find $\Gamma$ 
we need to integrate the anomaly equation 
$$\delta\Gamma={1\over 2}\left[(d+2)Tr{\delta\sigma}
e^{-\epsilon L^{\dagger}L}
-dTr{\delta\sigma}e^{-\epsilon LL^{\dagger}}\right]=
\int {\sqrt g}dx{\delta\sigma}{\it A}(g) , \eqno(27)$$
where ${\it A}(g)$ is the local conformal anomaly. 
To integrate the anomaly equation (27) we consider 
a one-parameter family of metrics $g(\tau)$, $\tau\in [0,1]$ 
interpolating between ${\bar g}$ and 
$g=e^{2\sigma}{\bar g}$: $g(\tau)=e^{2\tau\sigma}{\bar g}$. Then, 
${\delta\sigma}=d\tau\sigma$, ${\sqrt{g(\tau)}}=
e^{d\tau\sigma}{\sqrt{\bar g}}$. 
The local anomaly ${\it A}$ is given in terms of 
local curvature invariants for the metric $g(\tau)$. 
>From the transformation properties 
of the Riemann tensor under the local conformal rescaling 
we can easily find how the conformal anomaly 
${\it A}$ depends on $\tau$ and $\sigma$. 
The effective action $\Gamma$
is given by a simple integral over $\tau$
$$\Gamma[\sigma;{\bar g}]=
{\int_0}^1 d\tau\int dx{\sqrt{\bar g}}\sigma 
e^{d\tau\sigma}{\it A}[\tau,\sigma;{\bar g}] . 
\eqno(28)$$
Now we are in a position to express the ratio 
of two determinants evaluated at two metrics 
${\bar g}$ and $g=e^{2\sigma}{\bar g}$ in terms 
of the effective action $\Gamma$
$${\left[{det'L^{\dagger}L\over det<{\xi}_j|{\xi}_k>
det<h_j|h_k>}\right]^{1/2}_g=
e^{-\Gamma[\sigma;{\bar g}]}
\left[{det'{\bar L}^{\dagger}{\bar L}
\over det<{\bar\xi}_j|{\bar\xi}_k>
det<{\bar h}_j|{\bar h}_k>}\right]^{1/2}_{\bar g}}.
\eqno(29)$$
The dependence of the gravitational measure 
on the conformal factor $\sigma$ is obtained 
using eqs. (10) and (29)
$$\eqalign{d{\mu}(e^{2\sigma}{\bar g})=
e^{-\Gamma [\sigma;{\bar g}]}&
\left(det'{\bar L}^{\dagger}{\bar L}\right)^{1/2}_{\bar g}
\left[{det<\xi_j|\xi_k>_g det<h_j|h_k>_g\over 
det<{\bar\xi}_j|{\bar\xi}_k>_{\bar g}
det<{\bar h}_j|{\bar h}_k>_{\bar g}}\right]^{1/2}\times \cr 
&\times {det<\phi_j|h_k>_g\over det^{1/2}<h_j|h_k>_g}
{\it D}{\sigma}{\it D}'{\xi}{\it D}t \quad .\cr }
\eqno(30)$$
Using the conformal properties of $\phi_j$ and $h_j$ 
we finally arrive at the nicely factorized expression for 
the gravitational measure
$$d{\mu}(e^{2\sigma{\bar g}})=
e^{-\Gamma[\sigma;{\bar g}]}
\left({det'{\bar L}^{\dagger}{\bar L}\over 
det<{\bar\xi}_j|{\bar\xi}_k>}
\right)^{1/2}_{\bar g}{\it {dVol}}(Diff_0)
{\it D}{\sigma}d{\mu}_{\bar g}(t) , \eqno(31)$$
where 
$$d{\mu}_{\bar g}(t)=
{det<{\bar\phi}_j|{\bar h}_k>_{\bar g}\over 
det^{1/2}<{\bar h}_j|{\bar h}_k>_{\bar g}}{\it D}t , 
\eqno(32)$$
is the invariant, $\sigma$ independent, 
measure on the Schoen-Yamabe 
slice ${\bar Y}$, and ${\it {dVol}}(Diff_0)=
det^{1/2}<\xi_j|\xi_k>_g{\it D}'\xi$ 
is the measure on the group of difeomorphisms 
connected to identity $Diff_0$. 
The first factor in $dVol(Diff_0)$ is 
the volume form on the 
conformal group generated by CKV's $\xi_j\in KerL$. 
After dividing out the volume of the group 
of general covariance we are left with the 
measure on the Schoen-Yamabe slice $\bar Y$ 
and the induced effective action 
for the conformal factor $\sigma$. 
\par
The short-time heat kernel expansion 
for the vector and tensor operators 
$L^{\dagger}L$ and $LL^{\dagger}$ 
on manifolds without boundary contains 
local curvature invariants of an appropriate 
dimension (depending on $d$). The cases of interest 
for us are three- and four-geometries ($d=3$, $d=4$) 
for which we know the general form of the conformal anomaly. 
The only thing left is to determine 
exactly the numerical coefficients 
in front of different curvature invariants. 
In a separate paper we will 
discuss the exact form of 
a conformal anomaly for four-geometries. 
In the case of three-geometries the situation is 
much simpler. On a manifold without boundary the short-time 
heat hernel expansion has the universal form
$$Tre^{-t{\Delta}}=(4{\pi}t)^{-{d\over 2}}
\left(a_0+a_1t+a_2t^2+...
\right), \eqno(33)$$
where $a_k[g]$ are integrals of local curvature invariants. 
In particular, when $d$ is odd the conformal anomaly $a_{d\over 2}$ 
vanishes automatically. 
It can be present only when a manifold has a boundary, 
in which case the expansion (33) will have terms with half-integer 
powers of $t$. Those are the boundary terms which depend on the 
geometry of a boundary. In the case of three-manifold with a boundary 
the heat kernel expansion has a form
$$Tre^{-t{\Delta}}=(4{\pi}t)^{-{3\over 2}}\left(a_0+b_{1\over 2}
t^{1\over 2}+a_1t+b_{3\over 2}t^{3\over 2}+...\right), \eqno(34)$$
where $b_{1\over 2}$ and $b_{3\over 2}$ are the surface terms. 
$a_0$ is the cosmological (volume) term, $a_1$ is the Einstein-Hilbert 
term given by the integral of the Ricci scalar. In general, $b_{1\over 2}$ 
is the $2D$ cosmological term on the boundary and $b_{3\over 2}$ is the 
$2D$ Einstein term. The conformal anomaly in $d=3$ is given completely 
by the surface term. From (27) we can find the effective action 
which will have the volume and surface (boundary) contributions. 
The volume contribution, which is divergent, 
comes from integrating the Einstein-Hilbert action. Obviously, 
this anomaly (in any dimension) comes from the local action, and by 
adding a local counterterm which is the Einstein-Hilbert action we 
can cancel it. We conclude that the divergent volume part 
of the effective action $\Gamma$ renormalizes the cosmological 
and Newton constants. In other words, the Jacobian in the gravitational 
measure for three-geometries induces $3D$ gravity. 
By fine tuning the bare 
cosmological and Newton constants we can completely cancel the 
volume contribution to the effective action. What is left is the 
boundary contribution. But the boundary contribution to the conformal 
anomaly is the Einstein term, the Ricci scalar. Integrating this 
conformal anomaly we obtain the celebrated Liouville action on the 
boundary of a three-manifold. 
The total effective action $\Gamma$ has a form
$${\Gamma}={1\over 16{\pi}G}{\int d^{3}x(R-2{\Lambda})}+cI_L, 
\eqno(35)$$
where $I_L$ is the Liouville action depending on the metric 
on the boundary. We have seen that a quite general 
argument shows the presence of Liouville action in $3D$ gravity. 
In principle other terms involving the extrinsic curvature $K_{ij}$ 
could be present in the conformal anomaly. These are terms like 
$K_{ij}K^{ij}$, $K^2$, where $K={K_i}^i$, but by the Gauss-Coddazi 
equation only one of these terms could appear. However, such terms cannot 
appear because they would spoil the naive factorization of the measure. 
For any three-manifold with a boundary we can take the double of it 
and produce a manifold without a boundary. Now, the measure on 
a manifold without a boundary does not contain (surface) boundary 
contribution to the Jacobian. Naive factorization of the 
measure on the doubled manifold means that we have 
to take the product of two measures on a manifold with boundary and 
integrate over metrics on the dividing boundary. This requires the measure 
on a boundary which is the known measure for two-geometries. 
The measure over two-geometries has a Jacobian which is $exp(-26I_L)$, 
where $I_L$ is the Liouville action [2]. The dependence of the measure 
on a three-manifold with boundary on the geometry of a boundary must 
be of the same form, and cannot depend on the extrinsic curvature of 
a boundary. 
\par
We may ask a question, What is the meaning of the Liouville action 
in the measure for a three-manifold with a boundary ? 
Adopting the point of view of induced gravity, which proves to be 
quite useful for $d=2$ case, we can argue that by integrating over 
all metrics on a three-manifold with boundary with a fixed two-metric 
on a boundary (and fine tuning coefficients in the Einstein-Hilbert 
action to zero) we obtain the Hartle-Hawking wave function for the 
$3D$ Universe. The difference between $2D$ and $3D$ induced gravity 
is that in $d=2$ case the quantum measure induces the effective action, 
and in the $d=3$ case it is the Hartle-Hawking wave function which 
is effectively induced. Of course, one may argue that we have 
fine tuned coefficients in such a way that the Einstein-Hilbert action 
vanishes and we are not calculating the HH wave function for $3D$ 
gravity. Also, we get an infinite factor in the integral which is 
the volume of the space of metrics but this is the usual factor 
which can be taken care off by proper normalization of the HH wave 
function. We consider this interpretation of the appearance of 
the Liouville action in the induced HH wave function as a very 
attractive possibility. Indeed, it is well known since the work 
of Banks et al. [9] that 
the wave function for $3D$ gravity which is the solution of the 
Wheeler-DeWitt equation in the gaussian approximation is given 
in terms of the Liouville action evaluated to the same gaussian order. 
It is plaussible, therefore, that the exact HH wave function (not only 
the induced one) is given in terms of the Liouville action. 
\par
As a side remark we would like to describe 
how the induced HH wave 
function appears in the Chern-Simons gravity (or in any Chern-Simons 
$3D$ model with any gauge group). Witten has shown that on-shell, 
i.e., classically, the Einstein-Hilbert metric formulation of $3D$ 
gravity and the Chern-Simons gravity which is the $ISO(2,1)$ gauge 
theory are equivalent [10]. Now, the Chern-Simons gravity is 
(super)renormalizable, but naively the Einstein-Hilbert gravity is 
nonrenormalizable. The situation here is quite similar to that 
which occurs in string theory. In the Polyakov formulation the 
string model action is classically equivalent 
to the Nambu-Goto form of the string action. 
The Nambu-Goto action is nonrenormalizable but in the Polyakov 
formulation the model is renormalizable. 
The partition function for the Chern-Simons gravity is given in terms 
of the Ray-Singer analytic torsion invariant and does not depend on 
the $3D$ metric which was chosen to gauge fix the action. This happens 
because the contributions to the effective action, which may depend 
on the metric, from the ghosts and the gauge field cancel exactly. 
In any case, the possible (volume) contribution would have had the form 
of the Einstein-Hilbert action if it were not vanishing exactly. 
What happens when we consider the partition function of the Chern-Simons 
gauge theory for the three-manifold with a boundary ? 
In this case it is easy to convince ourselves that the dependence 
on the gauge-fixing metric is nontrivial because the boundary 
contributions to the conformal anomaly from ghosts and gauge fields 
do not cancel each other. In fact, the boundary contribution to 
the partition function is given 
by $exp(cI_L)$, where $c$ is the central charge of the corresponding 
conformal field theory which ``lives'' on the boundary. In this sense 
the topological character of the Chern-Simons gravity (and other CS 
models with different gauge groups) is broken by gravitational 
conformal anomaly. Boundaries are responsible for breaking the 
topological invariance of CS model. Once again we may adopt a point of 
view that in analogy to $2D$ case, where for each conformal field theory 
we get induced gravity, any Chern-Simons $3D$ gauge model leads to 
induced $3D$ gravity. The only difference is that in the $3D$ CS case 
we get the induced HH wave function of the Universe, rather than the 
induced action. 
\par
We conclude with a short discussion of the main results of this paper. 
The main new result appears to be the construction of the universal 
(for any dimension $d$) quantum gravitational measure which generalizes 
the Polyakov measure for two-geometries. The difference between odd 
and even dimensions was pointed out, and the special example of 
the quantum gravitational measure for three-geometries shows 
the important role of boundaries in odd dimensions. In particular, 
we argued that the metric dependence of the measure 
for three-geomerties is given by the Einstein-Hilbert action and 
the boundary contribution which 
is given in terms of the Liouville action. 
This quite mysterious fenomenon of appearance of the Liouville action 
in quantum $3D$ gravity, encountered earlier by Banks et al., 
is interpreted in terms of the induced HH wave function (rather than 
induced action). We compared this situation to that which occurs 
in the case of the Chern-Simons gravity where we get an exact induced 
HH wave function for $3D$ gravity as a result of conformal anomaly. 
The more complicated case of the measure for quantum four-geoemtries 
requires an exact calculation of the effective action for the conformal 
factor. This will be discussed in the forthcoming paper. 
\par
I would like to thank K. Aoki, E. D'Hoker, J. B. Hartle, E. Mottola, 
T. Tomboulis for discussions on the subject 
of the gravitational measure, R. Schoen for explaining 
to me some nuances of the Yamabe problem, and R. Peccei for 
continuing support. 
I also enjoyed discussion with T. Banks and L. Susskind 
about the WDW wave function in 3D gravity.
\endpage
\vskip 1cm
\centerline{\bf REFERENCES}	
\item{ 1)}  R. P. Feynman, Rev. Mod. Phys. {\bf 20} (1948) 367.
\item{   }  R. P. Feynman and A. R. Hibbs, Quantum mechanics and 
path integrals, 
(McGraw-Hill, New York 1965).         
\item{ 2)}  A. M. Polyakov, Phys. Lett. {\bf B103} (1981) 207, 211.    
\item{ 3)}  E. D'Hoker and D. H. Phong, Rev. Mod. Phys.{\bf 60} (1988) 917.    
\item{ 4)}  P. O. Mazur and E. Mottola, Nucl. Phys. {\bf B341} (1990) 187.    
\item{ 5)}  A. M. Polyakov, Mod. Phys. Lett. {\bf A2} (1987) 893.
\item{   }  V. G. Knizhnik, A. M. Polyakov and A. B. Zamolodchikov, 
Mod. Phys. Lett. {\bf A3} (1988) 819. 
\item{   }  J. Distler and H. Kawai, Nucl. Phys. {\bf B321} (1989) 509.     
\item{ 6)}  A. E. Fischer, Relativity: Proc. Relativity Conf. in the Midwest, 
\item{   }  ed. M. Carmeli, S. I. Fickler and L. Witten 
(Plenum, New York, 1970) pp. 303-357. 
\item{ 7)}  B. S. De Witt, Relativity, groups and topology 
(Gordon and Breach, New York, 1964).  
\item{   }  B. S. De Witt, Phys. Rev. {\bf 160} (1967) 1113.
\item{ 8)}  R. Schoen, J. Diff. Geom. {\bf 20} (1984) 479.    
\item{ 9)}  T. Banks, W. Fischler and L. Susskind, Nucl. Phys. {\bf B262} 
(1985) 159.     
\item{10)}  E. Witten, Nucl. Phys. {\bf B311} (1988) 46.    
\end